\documentclass[article]{svjour2}
\usepackage{graphicx}
\usepackage{rotating}
\usepackage{amssymb}
\usepackage{mathptmx}
\usepackage{color}
\usepackage[numbers]{natbib}
\journalname{Journal of Low Temperature Physics}
\bibpunct{}{}{,}{s}{}{,}

\usepackage{float}

\begin{document}

%\preprint{}

\newcommand{\hdblarrow}{H\makebox[0.9ex][l]{$\downdownarrows$}-}

\newcommand{\eqref}[1]{Eq.~\ref{#1}}
\newcommand{\eqsref}[2]{Eqs.~\ref{#1} and \ref{#2}}
\newcommand{\figref}[1]{Fig.~\ref{#1}}

\title{Weak-Link Phenomena in AC-Biased Transition Edge Sensors.} %Title of paper


%\author{L.Gottardi}
%\affiliation{SRON National Institute for Space Research,Sorbonnelaan 2, 3584 CA Utrecht, The Netherlands}

\author{L. Gottardi $^1$ \and H. Akamatsu \and M. Bruijn $^1$ \and J.-R. Gao$^1$,$^2$
  \and R. den Hartog $^1$ \and R. Hijmering $^1$ \and H. Hoevers $^1$ \and
P. Khosropanah $^1$ \and A. Kozorezov $^3$ \and J. van der Kuur $^1$
\and A. van der Linden $^1$ \and M. Ridder $^1$}

\institute{1:SRON National Institute for Space Research,\\
Sorbonnelaan 2, 3584 CA Utrecht, The Netherlands\\
%Tel.:\\ Fax:\\                                                                 
%\email{l.gottardi@sron.nl}                                                     
\\2: Kavli Institute of NanoScience, Faculty of Applied Sciences,\\
 Delft University of Technology,\\
Lorentzweg 1, 2628 CJ Delft, The Netherlands\\
\\3: Department of Physics, Lancaster University,\\
 LA1 4ER, Lancaster, UK
}

\date{\today}

\maketitle %\maketitle must follow title, authors, abstract and \pacs


\begin{abstract}
It  has  been recently  demonstrated  that superconducting  transition
edge-sensors behave as weak-links due to longitudinally induced
superconductivity  from the  leads with  higher $T_c$.   In this  work we
study the  implication of this behaviour for  TES-based bolometers and
microcalorimeter under ac bias.   The TESs are read-out at frequencies
between  1 and  $5 \mathrm{MHz}$  by a  Frequency Domain  Multiplexer based  on a
linearised two-stage SQUID  amplifier and high-$Q$ lithographically made
superconducting $LC$ resonators. In particular,  we focus on SRON TiAu
TES  bolometers  with  a  measured  dark  Noise
Equivalent Power    of  $3.2\times  10^{-19}
\mathrm{W/\sqrt{Hz}}$ developed for the  short wavelength band for the
instrument SAFARI on the SPICA telescope.

%\textcolor{red}{We report evidence of the Josephson effects observed in the TES
%current-to-voltage characteristics as well as in the sensitivity of the
%current  to  the  applied   magnetic  field.  An  explanation  of  the
%experimental results is given  within the standard resistively shunted
%junction (RSJ) model}.
\end{abstract}

\section{Introduction}
 \label{intro}
%\subsection{}
%\subsubsection{}

Superconducting transition-edge sensors (TESs) are highly sensitive
thermometers widely used as radiation detectors over an energy range from
near infrared to hard x-ray.
TESs can be operated both in the  dc and ac bias mode and in both
cases the detector response can be
modelled in  great detail \cite{Swetz12,JvdKuur11}. It has been recently demonstrated that TES-based devices behave
as weak-links due to the proximity effect from the superconducting leads \cite{Sadleir10}. A detailed experimental
investigation of the weak-link  effects in dc biased x-ray
microcalorimeters is ongoing \cite{Sadleir11,Smith12} and a
theoretical framework for modelling of the resistive state of a TES under dc bias has been
developed \cite{Kozorezov11}. Evidence of weak-link effects in ac biased
TES microcalorimeters where reported\cite{Gottardi12xray}, but
an adequated experimental and theretical investigation is still missing.
 In this report we  present a detailed characterization of a TES-based
 low-$G$ bolometer developed for the Short Wavelength Band detector of the instrument
 SAFARI on board of the Japanese infrared mission SPICA \cite{Safari}. A
 comparison between the performance of the TES under dc and
 ac bias is reported. 

\section{Experimental set-up}
 \label{Setup}

For the ac measurements described below we use a Frequency Domain
Multiplexer (FDM) system \cite{GottardiSPIE2012} working in the
frequency range from 1 to $5 \mathrm{MHz}$.
The read-out is done using a two-stage
SQUID amplifier with on-chip linearization from PTB and high-$Q$
lithographic $LC$ resonators \cite{Gottardi12bolo}. 
The TES arrays chip is mounted on a copper bracket inside a
superconducting Helmholtz coil used to generate a
uniform perpendicular magnetic field over the whole pixels array. 
The external magnetic field  is shielded using a Nb can covered by few layers of metallic glass  tape. 

The device under test is a low-$G$ bolometer based on a Ti/Au (16/60 nm)
bilayer, deposited on a $0.5 \mu \mathrm{m}$ thick, $130\times 70 \mu
\mathrm{m}^2$ suspended $Si_3N_4$ island connected to the thermal bath via
4 $Si_3N_4$ cross-shaped supporting legs that are $2 \mathrm{\mu m}$ wide and $400 \mathrm{\mu m}$ long.
The TES area is $50\times 50 \mu \mathrm{m^2}$.  It has a critical
temperature of $T_C=85\,\mathrm{mK}$, a  normal state  resistance  of
$R_N=98\,\mathrm{m\Omega}$, a measured $G=0.27 \mathrm{pW/K}$ and a calculated
Noise Equivalent Power (NEP) of $2.\
3\times 10^{-19} \mathrm{W/\sqrt{Hz}}$.
An $8\mathrm{nm}$ thick Ta absorber with an area of $70\times70 \mathrm{\mu m}^2$ is deposited close to the TES. 

\begin{figure}[htbp]
    \centering
    \includegraphics[height=5.2cm]{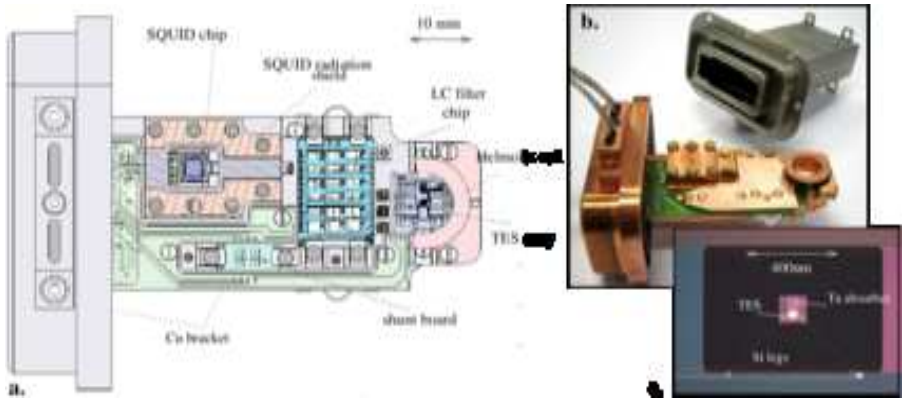}
    \caption[example] 
%>>>> use \label inside caption to get Fig. number with \ref{}
   { \label{fig:FDMscheme} The schematic drawing ({\bf a.}) of the
     Frequency Domain Multiplexer and pictures of the experimental set-up ({\bf b.})  and the TES bolometer with cross-shaped supporting legs ({\bf c.}).}
   \end{figure}

The electrical contact to the bolometer is realized by $90\mathrm{nm}$ thick
Nb leads deposited on the top of the SiN legs.
The sensor was previously characterised
under dc bias \cite{Pourya12,Hijmering12}  and showed a power plateau of
$9.4\mathrm{fW}$ and a dark NEP of $4.8\times 10^{-19}
\mathrm{W/\sqrt{Hz}}$ at $30\mathrm{mK}$.
Below we report the results for the TES ac biased at a frequency of
$2.4\mathrm{MHz}$.
The FDM set up and a picture of the TES bolometer is shown in Fig.~\ref{fig:FDMscheme}.

\section{Experimental results}
 \label{results}

\begin{figure}[ht]
\center
\includegraphics[height=4.7cm]{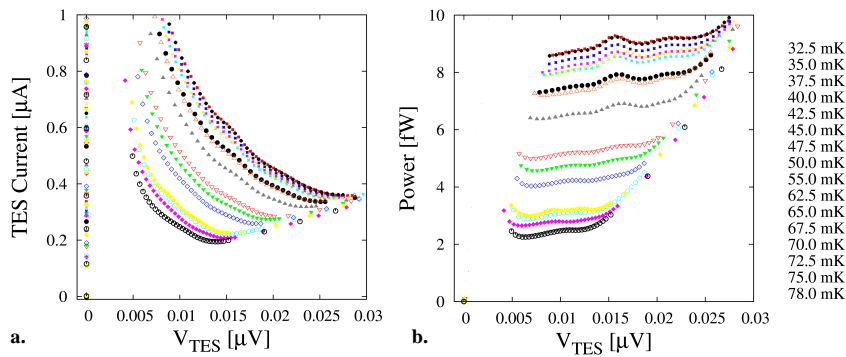}
\caption{\label{fig:ivpv} Measured TES current ({\bf a.}) and
  power ({\bf b.}) as a function of the ac bias voltage for
  several bath temperature}
\end{figure}

To characterize the detector under ac bias we studied the dependence of
the TES current on the voltage, the bath temperature and the applied
magnetic field.
In \figref{fig:ivpv} we show the TES current-to-voltage ($IV$) and
power-to-voltage ($PV$) 
characteristics as a function of the bath temperature. 
The TES current shows regular oscillating structures that are better
resolved when looking at the measured power. The maxima of those
structures occur at relatively constant TES voltages for all the bath
temperatures.

\begin{figure}[ht]
\center
\includegraphics[height=8cm,angle=-90]{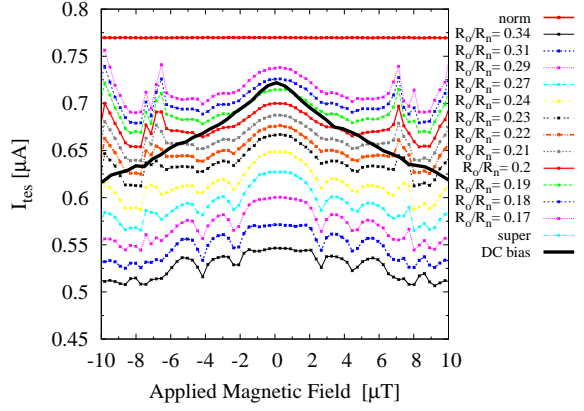}
\caption{\label{fig:IvsB} Measured TES current as a function of
  applied magnetic field.}
\end{figure}

The dependence of the TES current on the applied magnetic field is
shown in \figref{fig:IvsB}. Under ac bias (colored-dotted lines) with
the TES in transition  we
observed a Fraunhofer-like oscillating pattern, typical of a superconducting
weak-link structure. 
Under dc bias (dark line) the current oscillations are generally much
smaller and more difficult to see as it was  previously reported
\cite{Hijmering12}.  Another interesting difference between the ac and dc
bias measurement is that at large applied  magnetic field the ac
current does not decrease as one would expect from the Meissner
effect. Moreover, a sudden increase of the TES current is observed at
certain  magnetic fields.  

Despite the observed structure in the $IV$s characteristics  
the TES bolometer dark NEP at zero magnetic field was measured to be
only a factor 1.4 higher than the theoretical value. 

\begin{figure}[ht]
\center
\includegraphics[height=7.5cm,angle=-90]{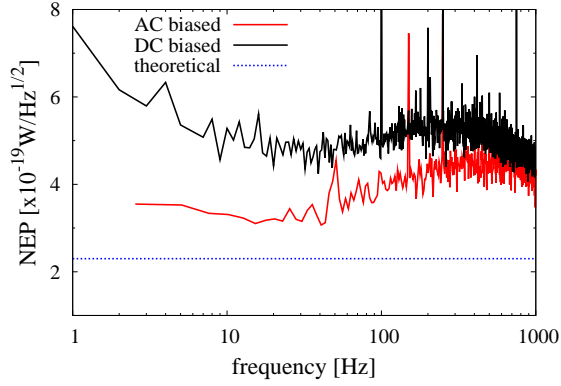}
\caption{\label{fig:nepacdc} Measured dark NEP at low frequency under
  ac and dc bias. The spectra were taken at TES resistance
  $R_{tes}=0.3R_{N}$ and at a bath temperature of 
$T_{bath}=40 \mathrm{mK}$ and $T_{bath}=20 \mathrm{mK}$
  respectively. In the ac bias case a double light-tight box was used.}
\end{figure}

In \figref{fig:nepacdc} the dark NEP 
spectra taken under ac and dc bias  at a TES resistance
$R_{tes}=0.3R_{N}$  and a 
bath temperature of respectively
$T_{bath}=40 \mathrm{mK}$ and $T_{bath}=20 \mathrm{mK}$
are shown. The dark NEP was calculated by dividing the
current noise by the responsivity at low frequency, which can be
approximated by $\frac{1}{I_O(R_{tes}-Z_{th})}$, where $I_O$ is the
effective bias current, $R_{tes}$ is the TES resistance and $Z_{th}$
is the Thevenin impedance in the bias circuit as derived from the
calibration of the $IV$ curves.

The dark NEP measured at low frequency( $f \lesssim 40 \mathrm{Hz}$) at
the optimal bias point in the transition and at zero magnetic field was about $(3.2\pm 0.13)\times 10^{-19} \mathrm{W/\sqrt{Hz}}$
 and $(4.8 \pm 0.2)\times 10^{-19} \mathrm{W/\sqrt{Hz}}$ for
the ac and dc bias case respectively. 
The NEP was independent on the bath temperature for temperature
approximately below $50\mathrm{mK}$.
The noise measured under ac bias is only 
1.4 times higher than the expected theoretical NEP of  $2.3\times
10^{-19} \mathrm{W/\sqrt{Hz}}$ and was obtained after mounting the
light-tight FDM set-up in a second light-absorbing box.
We suspect that even this approach was not sufficient to reduce the effect of the stray radiation to below $10^{-19} \mathrm{W/\sqrt{Hz}}$ and a further improvement of our experimental set-up is needed.
 
We remark that the measurements under dc bias were done using  a
single light-tight box. They are affected  at low frequency by an
excess of power leaking throught the box and by the $1/f$-noise from the SQUID
amplifier.

\begin{figure*}[ht]
\includegraphics[width=11.5cm]{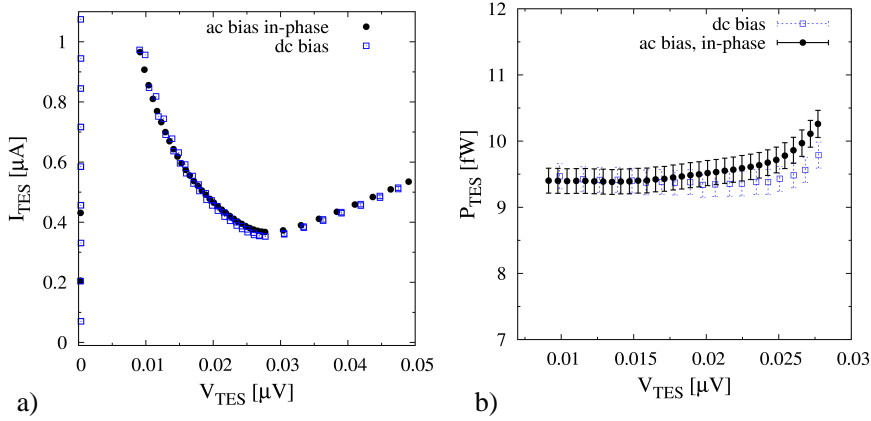}
\caption{\label{fig:pivacdc} $IV$ ({\bf a}) and $PV$ ({\bf b})
  characteristics of the TES-based bolometer measured under ac and dc bias.}
\end{figure*}

The observed TES current response reported above are
  likely due to the weak-link behaviour of the bolometer. A detailed
  explanation of the Josephson effects in an ac-biased TES will be reported in a
following paper. It can be shown that in order to compare the performance of the TES under ac and dc bias, only the current in-phase with the applied voltage should be considered
 in the ac bias case.  The rms value of the in-phase ac current is equivalent to the dc current value  measured in a dc
 biased TES. This has been experimentally verified and the results are shown in
 Fig~\ref{fig:pivacdc}.a. 
The current-to-voltage  characteristics measured under ac  and dc bias
show good  agreement within the experimental uncertainties  due to the
calibration  procedure.   Under  ac  bias  the  quasiparticle  current
$I_{qp}$, being in-phase with the applied voltage, is the only current
which contributes to  the power dissipated in the  TES.  The data sets
plotted   in    Fig~\ref{fig:pivacdc}.b   are   the    measured   $PV$
characteristics of  the bolometer under  ac and dc  bias respectively.
The $PV$ curves are consistent  with each other within the calibration
errors.  We clarify here that  the TES current is calibrated using the
nominal SQUID  parameters and using standard  SQUID signal calibration
techniques, while to obtain the TES voltage we  assumed the  TES normal state  resistance $R_N$ to  be known  and  performed a
linear fit of the superconducting  and normal branches of the measured
$IV$ curves. The TES $R_N$ has been measured on test TiAu  structures  with  4  points  measurements in  a  dedicated  test
set-up. The error on $R_N$ is about $1\%$.  This calibration procedure
is particularly  sensitive to  how accurate the  normal branch  of the
$IV$ curve  has been  measured.  
%, in  particular, under  dc-bias, it
%requires  a very  precise estimation  of the  offset around  zero. The
%ac-bias  measurements are  intrinsically  not affected  by the  offset
%calibration.  
In  both the ac and  dc bias case we  estimate the total
uncertainties to  be about  $2\%$ and $4\%$  in the current  and power
respectively.

\section{Conclusion}

We observed weak-link behaviour in the ac biased  TES-based low G
bolometer developed for SAFARI. A complete modeling of the
weak-link effect in an ac biased detector is under development. When
looking only at the current component in-phase with the applied
bias voltage the TES current and power response   under ac bias is
comparable with the dc data. 
 A dark NEP of $(3.2\pm 0.13) \cdot 10^{-19}                             
\mathrm{W/\sqrt{Hz}}$ was observed with the pixel ac-biased  at $2.4
\mathrm{MHz}$. The measured dark NEP is only a factor of $1.4$ higher than the
expected theoretical value and is very likely affected by a not yet stray-light free environment.

% If in two-column mode, this environment will change to single-column format so that long equations can be displayed. 
%\begin{widetext}
%$$\mbox{put long equation here}$$
%\end{widetext}

% Figures should be put into the text as floats. 
% Use the graphics or graphicx packages (distributed with LaTeX2e).
% Put the label that you will use with \ref{} command in the braces of the \label{} command.
%
% \begin{figure}
% \end{figure}

% \begin{table}
% \end{tabular}
% \end{table}

%\begin{acknowledgments}
% Put your acknowledgments here.
%\end{acknowledgments}

\bibliography{Gottardi_LTD15}
\bibliographystyle{unsrt}   
\end{document}